# A Generalised Nucleation Theory for Ice Crystallization

Maodong Li[1], Yupeng Huang[2], Yijie Xia[2], Dechin Chen[1], Cheng Fan[2], Lijiang Yang[2], Yi Qin Gao[1,2*], Yi Isaac Yang[1*]

[1]Institute of Systems and Physical Biology, Shenzhen Bay Laboratory, Shenzhen 518132, China

[2]College of Chemistry and Molecular Engineering, Peking University, Beijing 100871, China

*Corresponding authors: Yi Isaac Yang (yangyi@szbl.ac.cn), Yi Qin Gao (gaoyq@pku.edu.cn)

**Abstract**

Despite the apparent simplicity of water molecules, the kinetics of ice nucleation under natural conditions can be surprisingly intricate. Previous studies have yielded critical nucleation sizes that vary widely due to differences in experimental and computational approaches. In our investigation, we employed all-atom molecular dynamics simulations to explore spontaneously grown and ideal ice nuclei, revealing significant disparities in their kinetics. Notably, nucleation defects challenge the applicability of classical nucleation theory (CNT) to spontaneously grown ice nuclei. To address this, we propose a generalised nucleation theory that effectively describes the kinetics of ice crystal nucleation across diverse conditions. The kinetics of ice nuclei, as characterised by the "corrected" critical nucleus size, follow a linear law akin to that assumed by CNT. This generalised nucleation theory also provides insights for studying the kinetics of other crystalline materials.

## 1 Introducion

Water, a seemingly simple substance, exhibits remarkable complexity with over eighteen distinct phases[1,2]. When cooled at atmospheric pressure, water's thermodynamic equilibrium state is hexagonal ice (ice $I_h$)[3], while cubic ice (ice $I_c$)[1,3,4] represents a meta-stable phase[5]. Experiments were performed to obtain pure ice $I_c$[6-8], and only recently was defect-free ice $I_c$ obtained[1]. Molecular dynamics (MD) simulations offer an atomic-level perspective on phase transitions, including nucleation processes[9-12]. In particular, enhanced sampling techniques[13] enabled the study of both homogeneous nucleation of pure ice $I_c$ or pure ice $I_h$ through these simulations[14,15].

However, the well-known Ostwald's step rule[16] assumes that ice first nucleates in the form of ice Ic and then transforms into ice Ih with a long relaxation time[17,18], which hints that ice $I_c$ is faster in kinetics. Classical nucleation theory (CNT) is widely used to describe the kinetics of homogeneous nucleation. As a specific choice of the dividing surface under the incompressibility approximation, the formation of a critical nucleus ($N_c$) requires overcoming a free-energy barrier[19-21], $\Delta G_c$:

$$\Delta G_C = \frac{16\pi\gamma^3}{3\rho_s^2 |\Delta\mu|^2}, \tag{1}$$

where $\rho_s$ is the density of the solid phase, $\Delta\mu$ is the chemical potential difference between the solid and the fluid phase at the same temperature and pressure at which the cluster is critical, and $\gamma$ is the surface free energy. When we assume a spherical shape for a defect-free cluster, $N_c$ can be estimated by[19,22]:

$$N_C = \frac{32\pi\gamma^3}{3\rho_s^2 |\Delta\mu|^3}. \tag{2}$$

Critical ice nuclei are relatively small and short-living[18,19], making them difficult to observe in experiments. Nc is a temperature-sensitive value and becomes infinity at the melting temperature $T_m$[19-21]. Therefore, many researchers have investigated supercooled water around the so-called "no man's land" temperature region (230 K)[14,18,19,21,23,24]. Surprisingly, the values of $N_c$ obtained in different studies varied considerably, ranging from 100 to 600 (Supplementary Table S1). The uncertainty in the measured or calculated values of $N_c$ suggests that CNT (Eq. (1-2)) is very rudimentary for describing the kinetics of ice nucleation in the natural state.

In this paper, we employed all-atom MD simulations to explore the kinetic differences between ideal and spontaneously grown ice crystal nuclei. A subset of the ice nuclei was grown within MD simulations using enhanced sampling methods, while another subset was spherically cut from ideal crystals. Remarkably, we observed significant disparities in the kinetics of these two types of ice clusters, which cannot be explained by CNT. Therefore, we proposed a modified nucleation theory to elucidate the kinetics of ice crystallisation under general conditions.

## 2 Materials and Methods
### 2.1 XRD3D collective variable

We first grew ice crystal nuclei in MD simulations. We have successfully obtained all possible ice states of the system in all-atom MD simulation[15] using the MetaITS enhanced sampling method[25] with the collective variables (CVs) based on X-ray diffraction (XRD) intensity. The CVs based on three-dimensional ($s_{3D}$) and two-dimensional ($s_{2D}$) XRD intensities are defined as follow:

$$s_{3D}(\mathbf{R}) = \frac{1}{N}\sum_{i=1}^{N}\sum_{j=1}^{N} f_i(Q) f_j(Q) \frac{\sin(Q \cdot R_{ij})}{Q \cdot R_{ij}} \cdot \omega(R_{ij}), \tag{3}$$

$$s_{2D}(\mathbf{R}) = \frac{1}{N}\sum_{i=1}^{N}\sum_{j=1}^{N} f_i(Q) f_j(Q) J_0(Q \cdot R_{ij}^{xy}) \cdot \omega^{xy}(R_{ij}^{xy}) \omega^{z}(R_{ij}^{z}), \tag{4}$$

where $f_i(Q)$ and $f_j(Q)$ are the atomic scattering form factors, $R_{ij}$ is the distance between atoms $i$ and $j$, $\omega(R_{ij})$ is the window function, and $J_0(Q \cdot R_{ij}^{xy})$ is the Bessel function.

In this study, we modified the CVs to control the region of phase transition, allowing water molecules to grow into ice clusters of an arbitrary size in MD simulations. To achieve better isotropy in homogeneous ice nucleation, it is strongly recommended adopting a three-dimensional spherical crystal nucleus as the initial configuration[19-22,26], thereby mitigating the influence of sharp crystal interfaces. We confined the calculation of the XRD intensity to a spherical shape by introducing a Fermi switching functions to the atomic scattering form factors $f_i(Q)$:

$$f'_i(Q) = f_i(Q) \cdot \frac{1}{1+\exp^{a(\|R_i-R_0\|_2 - r_{cut})}}, \qquad (5)$$

where $R_i$ and $R_0$ are the coordinate positions of atom i and a fix sphere centre, $r_{cut}$ is the maximum radius of the grown ice clusters and α is the buffer depth. Here we set α as 0.5 nm$^{-1}$ and $r_{cut}$ as 2.0 nm, a space that can accommodate up to 1060 ice molecules. Such spontaneously grown ice clusters inevitably contain internal defects, which we use to resemble ice nuclei through spontaneous formation (Fig. 1).

**2.2 Simulation Details**

All MD simulations were performed using the SPONGE[27] software package version 1.2 (https://spongemm.cn/en/) (Supplementary Fig.S1). The simulation systems contain 23040 transferable intermolecular potential with 4 points/ice (TIP4P/ice) water molecules in a periodic box with initial dimensions of 9.3×8.9×8.8 nm$^3$. We adjusted the temperature of the system to 230 K by Nose–Hoover thermostat temperature coupling with a relaxation time constants of 1.0 ps and the pressure of the system to 1 bar by Andersen barostat with a relaxation time constants of 2.0 ps.

**Sampling** We firstly performed multiple short-time MD simulations using the metadynamics[28] enhanced sampling methods to grow thousands of ice crystal nuclei with different sizes and polymorphisms. We performed 0.8 μs configurational sampling in the NPT ensemble, starting with a liquid conformation. We collected conformations every 1 ps for crystal structure identification (Supplementary Fig.S2-3). And we randomly selected ice nucleus seeds according to a same $P(N_{cluster})$ distribution (Supplementary Fig.S4).

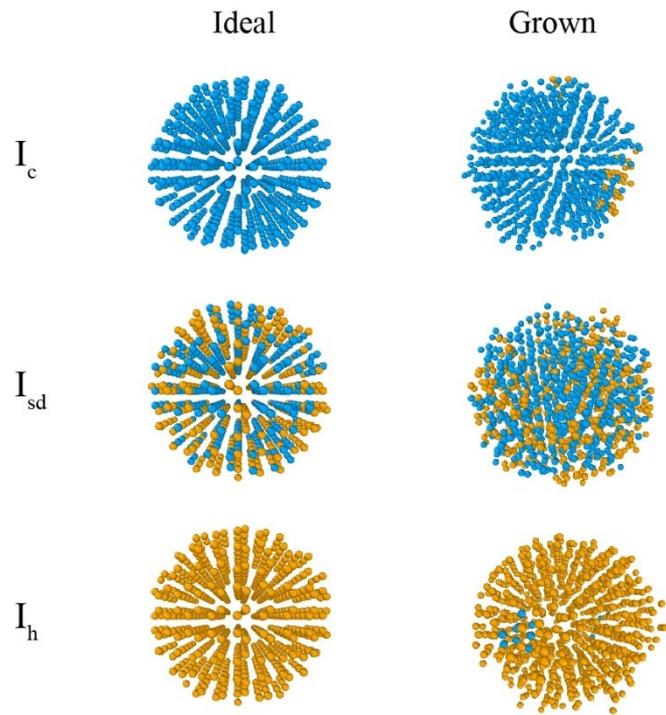

FIG. 1. Typical structures of ice nuclei. Each sphere represents an oxygen atom in a water molecule, where the blue and orange spheres represent water molecules in the cubic and hexagonal crystal states, respectively. (Left) Ideal ice nuclei are cut spherically from perfect crystals. (Right) Spontaneously grown ice nuclei nucleus by MD simulations.

**Heating-annealing** Secondly, to ensure each largest cluster is decoupling from the simulation box, we restrained the largest cluster of the box and heated it to 270 K for 1.0 ns, then cooled it to 230 K for 1.0 ns in the NVT ensemble.

**Prepare for the ideal group** For comparison, we also prepared a series of ideal spherical ice nuclei of different sizes and polymorphisms, which were cut from three ideal ice $I_c$, $I_{sd}$, and $I_h$ crystals[19]. The reference clusters with a radius of 0.7-1.9 nm are prepared from the ideal crystal conformation. These clusters can hold 50-1200 ice molecules. The ideal group also follows the process heating-annealing-seeding. Sixty samples were cut from each crystal giving a total number of 180.

**Seeding** Finally, we follow the "seeding" computational approach[19], which is widely adopted in nucleation studies. This scheme consists in introducing the cluster seed into a bulk of supercooled liquid and let it evolve in the NPT ensemble. In this large simulation system, the melting committor can be similarly defined when $N_{cluster}$ < 20, but the freezing destination is too hard to reach, which takes longer than microsecond to freeze completely in the box, as mentioned in the following section. Luckily, the shooting fate of most nuclei is determined in a short time. In all-atom simulations practice, a fixed shooting time is adopted (in references, 3.0 ns[19], 0.5 ns[22]). Here we performed 100-ns shooting at 230 K in the NPT ensemble (Supplementary Fig.S5). At the size of $N_c$, the crystal seed can grow or melt with equal probability.

### 2.3 Water-ice definition

The water-ice definition is given according to the refenences[15,29] (Supplementary Eq. (S1-S3)). In water crystal structure identification, only atom O positions are considered. Here we use the same definition as "identify diamond structure" in Ovito[29] and applied it into a new software SPONGE. To classify the type of a central molecule, this structure identification method considers the second nearest neighbours to discriminate between cubic and hexagonal diamond structures. Under this local definition, each molecule is defined into $N_c(I_c)$, $N_h(I_h)$, or $N_{liquid}$. Both Ice $I_c$ and Ice $I_h$ are defined into solid ice $N_{ice}$.

## 3 Results
### 3.1 Nucleation Kinetics

In the 100-ns aimless shooting, we define the size changes of nuclei $\Delta N$, to describe the shooting fate as follows. Fig. 2 shows the size change $\Delta N$ for different ice cluster size $N$ at the end of the "seeding" simulation. The "seeding" results indicate that CNT can describe well the kinetics of ideal ice nuclei. Fig. 2(a-c) shows that the variation of $\Delta N$ for the three ideal ice nuclei is in a positive linear relation with the ice cluster size. Therefore, we can easily determine their critical nucleus size $N_c$ that fits the definition of CNT, i.e., the number of water molecules contained in the ice nucleus when $\Delta N = 0$. The $N_c$ of the ideal ice nuclei of $I_c$, $I_{sd}$, and $I_h$ all lie within a range of 255 to 295. For these perfect nuclei without defects, there does not seem to be much difference among different polymorphisms.

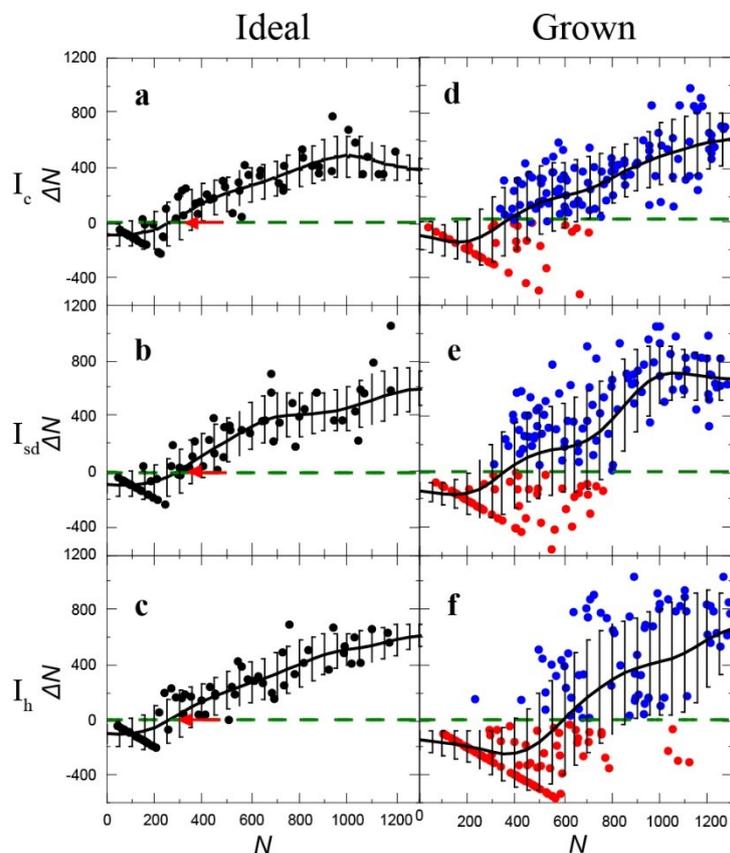

FIG. 2. Size changes $\Delta N$ of different ice polymorphs at the end of the "seeding" simulations versus their initial sizes $N$. From top to bottom are Ice $I_c$, Ice $I_{sd}$ and Ice $I_h$, respectively. The black line represents the average size change for each cluster size. The green dashed line marks $\Delta N = 0$, i.e. the growing-melting equilibrium line. (Left): Ideal ice nuclei. The red arrows indicate the location of the critical nucleus size $N_c$. (Right): Spontaneously grown ice nuclei. The black line represents the average size changes. The growing and melting data points are distinguished by blue and red colours, respectively.

However, spontaneously grown ice nuclei possess kinetics that is obviously different from those of ideal nuclei. Fig. 2(d-f) reveals that while the $\Delta N$ of the grown nuclei generally increases with increasing ice cluster size, the data points are highly scattered and do not show a linear relationship as in the case of ideal nuclei. To further investigate the differences between growing and melting ice nuclei, we distinguish these data points using different colours (blue and red, respectively) in the figures. It can be seen that even some large ice crystal nuclei with water molecule numbers greater than 800 are subject to melting. This is particularly evident for $I_h$ which we will discuss later with respect to Ostwald's step rule. In contrast, a number of small ice crystal nuclei with only about 200 water molecules do grow. This observation is in line with the vast variations in critical nucleus sizes obtained in different studies.

In contrast to ideal ice nuclei, the kinetics of spontaneously grown ice nuclei with varying polymorphisms exhibit notable differences. We computed several kinetics- related properties for these nuclei based on the CNT[19,22], as summarised in Table 1. Notably, the critical nucleation size $N_c$ of ice $I_{sd}$ is 351, a value closely aligns with thermodynamic calculations[14]. Our calculation results reveal that the growth rates of different polymorphic ice nuclei follow the order: Ice $I_{sd}$ ≈ Ice $I_c$ > Ice $I_h$. However, from a thermodynamic perspective[15], their stability ranking order is Ice $I_{sd}$ > Ice $I_h$ > Ice $I_c$ . To understand this discrepancy, we performed MD simulations for each of the three distinct polymorphic ice nuclei, extending the simulations up to 3 μs. Remarkably, after 2 μs of simulation, all three systems—comprising 23,040 water molecules each—essentially completely froze, eventually converging into a mixed Ice $I_{sd}$ state (Supplementary Fig.S5-6). This observation reaffirms the entropic preference of ice $I_{sd}$ during the process of crystallization.

Table 1 | Theoretical estimated nucleation rate of the grown nuclei.

|  | $I_c$ | $I_{sd}$ | $I_h$ | $I_{sd}$[[14]] |
|---|---|---|---|---|
| $N_c$ | 362 | 351 | 599 | 314 |
| $f^+$(/s) | 4.24E+11 | 6.32E+11 | 7.69E+11 | 1.9E+11 |
| $Z$ | 0.00705 | 0.00716 | 0.00548 | 0.0076 |
| $\gamma$ | 70.33 | 69.61 | 83.19 | 67.0 |
| $\Delta G(k_B T)$ | 61.41 | 59.55 | 101.62 | 52.8 |
| $\log_{10}(J/m^3/s)$ | 11.00 | 11.99 | -6.31 | 14.8 |

The supercooling temperature $\Delta T$ = 40 K, $\Delta\mu$ = 0.155 kcal/mol, as Ref. [[14]]. $f^+$ is the attachment rate of particles to the critical cluster. $Z$ is the Zeldovich factor. $\gamma$ is liquid−solid surface free energy. $\Delta G c$ is nucleation free-energy barrier height. $J$ is the nucleation rate.

### 3.2 Nucleus Defects

We speculate that the difference between spontaneously grown and ideal ice nuclei is related to the defects present in the former but not the latter. Since the ideal nuclei are defect-free spherical nuclei cut from perfect ice crystals, they grow isotropically. In contrast, spontaneously grown nuclei are characterised by a variety of defect structures, such as 5-8 water ring[30], five-fold twin boundaries[18], and predominant stacking in more than one direction[23]. These defects are expected to affect the kinetics of nuclei growth, causing them to differ from the ideal nuclei.

We first examined the surface defects of ice nuclei. Since different from the ideal nuclei, the grown nuclei are most likely not perfectly spherical, their surface area thus differ from that of the ideal nuclei with the same cluster size. We used the solvent accessible surface areas (SASA) $S$ to denote the surface area of the nucleus and found that the ideal nucleus indeed has the smallest surface area for the same number of cluster particles, which is essentially proportional to $N^{2/3}$ (Supplementary Fig.S7). In contrast, spontaneously growing nuclei tend to have a larger surface area, suggesting a complex surface morphology. However, the surface area alone was shown unable to distinguish whether the nuclei will eventually grow or melt. Fig. 3(a) illustrates the relationship between $\Delta N$ and $S$, demonstrating that the final fate of spontaneously growing nuclei can be very different even when they have the same surface areas.

We next investigated the impact of internal defects of ice clusters on nucleation kinetics. We focused on the orientational order parameter $q$[14,31,32], a metric for assessing ice nucleus defects. Specifically, $q$ characterises tetrahedral configurations and reflects the lattice perfection of crystals with a coordination number of four, such as ice[14]:

$$q = 1 - \frac{3}{8} \sum_{\alpha > \beta} \left( \cos\theta_{\alpha\beta} + \frac{1}{3} \right)^2, \tag{6}$$

where $\theta_{\alpha\beta}$ represents the angle formed by the lines connecting the oxygen atom of a given water molecule to those of its 4 nearest neighbours, denoted as $\alpha$ and $\beta$. Fig. 3(b) demonstrates a strong positive correlation between $\Delta N$ and $q$ for both ideal ice nuclei and spontaneously grown ice nuclei. Furthermore, this parameter effectively distinguishes the growth behaviours of the two ice nucleus types, suggesting that $q$ serves as a valuable indicator for ice nucleus kinetics.

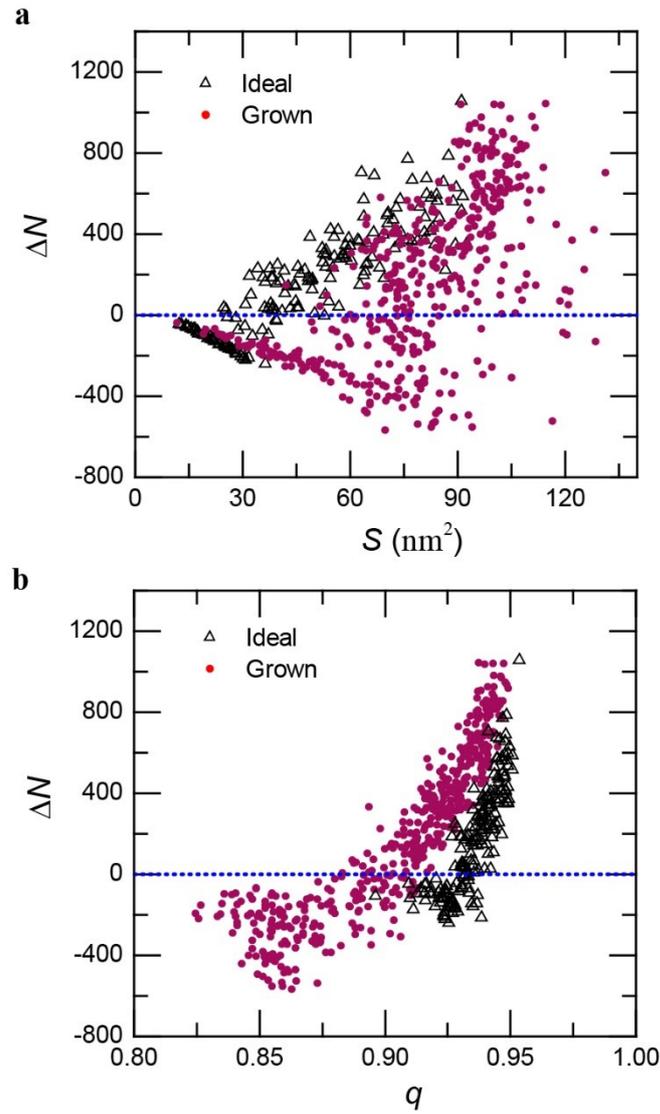

FIG. 3. Influence of internal and external defects in ice clusters on nucleation kinetics. Hollow black triangles indicate ideal ice nuclei and solid purple circles indicate spontaneously grown ice nuclei. The blue dashed line marks the position where $\Delta N = 0$. Plot of (a) ice cluster size changes $\Delta N$ versus solvent accessible surface areas (SASA) $S$, (b) Ice cluster size changes $\Delta N$ versus orientation order parameter $q$.

Consequently, incorporating the orientation order parameter $q$ alongside the nucleus size $N$ provides a more comprehensive description of ice nucleation kinetics (see Fig. 4(a)). Notably, Fig. 4(a) reveals a discernible boundary between growing and melting ice nuclei, which applies to both ideal and spontaneously grown nuclei. However, the impact of $q$ on ideal ice nuclei is negligible, given their minimal variation (all exhibiting large values). In contrast, for spontaneously grown ice nuclei, the order parameter $q$ is of similar importance as the nucleus size $N$ in influencing kinetics. Even when the ice nucleus size $N$ is large, it can eventually melt if the order parameter $q$ is small—indicating a high prevalence of internal defects. Thus, the directional order parameter $q$, characterising these internal defects, serves as an effective complement to the CNT approach, providing one more dimension for our understanding of natural ice nucleation kinetics.

### 3.3 Nucleus Defects

The orientation order parameter $q$ can also serve as a correction factor for the ice nucleus size $N'$, making it compatible with the conventional CNT approach. Here, we propose an effective ice nucleus size $N'$ that is "corrected" by the following relation:

$$N' = N\left(1 - M(q_{max} - q)\right), \tag{7}$$

where $M$ represents the second-order neighbour number of the crystal structure, and $q_{max}$ is the fitting coefficient for ideal nuclei. For ice nuclei, we have $M = 12$ and $q_{max} = 0.95$. The relationship between the "corrected" effective ice nucleus sizes $N'$ and the size changes $\Delta N$ is depicted in Fig. 4(b). Notably, both ideal and spontaneously grown ice crystal nuclei exhibit a strong positive correlation with the effective size $N'_c$ concerning their size change $\Delta N$. Furthermore, there is no significant difference in the kinetic behaviours between the two types of ice nuclei when they are described using the effective size $N'$. Specifically, the data points for spontaneously grown ice nuclei at $\Delta N = 0$ are no longer as scattered as shown in Fig. 2(d-f). We have determined $N'$ values for different polymorphic ice nuclei, which fall within a narrow range of 187 to 192. Using Eq. (7), both ideal and spontaneously grown ice crystal nuclei exhibit a strong positive correlation with the effective size $N'$ among various polymorphisms (Supplementary Figure. S8-9,

Table.S2).

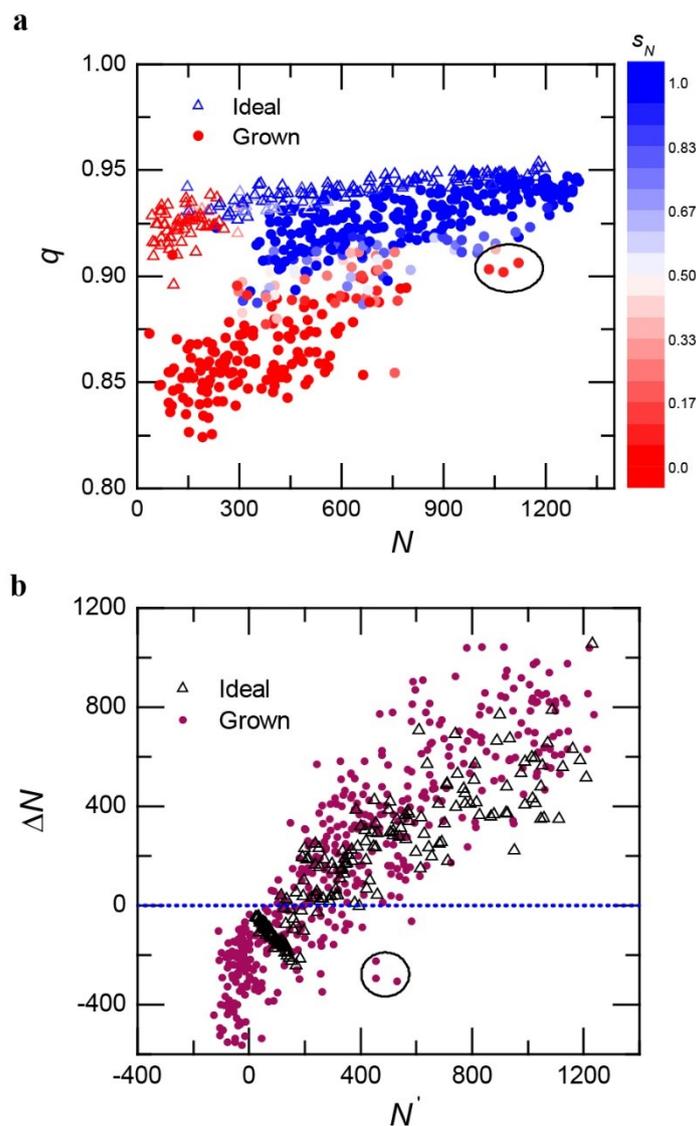

FIG. 4. The crystallisation kinetics versus various cluster descriptors. Hollow triangles indicate ideal ice nuclei and solid circles indicate spontaneously grown ice nuclei. The black circle marks anomalous data points. (a) Scatter plot of the normalised relative ratios of size changes $S_N$ as a function of the ice nucleus size $N$ and the orientation order parameter $q$. Data points are colour mapped by the value of $S_N$, with blue representing growth and red melting. (b) Size changes $\Delta N$ of ice nuclei versus "corrected" effective nuclei sizes $N'$. The blue dotted line marks the position where $\Delta N = 0$.

However, we do observe three anomalous data points in Fig. 4(a). These data points correspond to large nucleus sizes $N'$, and their order parameter $q$ values are also relatively large. However, their size changes $\Delta N$ are negative. These three points show a behaviour quite distinct from the other data points in the effective nucleus size $N'$ illustrated in Fig. 4(b). Upon examining the structure of these three ice clusters, we discovered that they exhibit poly-crystalline characteristics. Specifically, their main bodies consist of $I_h$ crystal structures coexisting with $I_c$ fragments. During the "seeding" process, the coexisting $I_c$ fragments rapidly dissolved or melted. Consequently, by the end of the 100 ns molecular MD simulation, the total number of molecules in the cluster showed a decrease from its initial value. However, the remaining portions of these ice clusters continued to grow slowly, as evidenced by our extension of the MD simulations for these three ice nuclei over a longer time (Supplementary Fig.S10-11). After examining the initial crystal structures of all other ice nuclei, we confirmed that only these three $I_h$ ice nuclei exhibit such distinct poly-crystalline features. This observation suggests that these three cases are indeed special occurrences. However, it raises an intriguing question: Why only do the $I_h$ ice nuclei have fragments of other crystals adhering to them? The answer might lie in the Ostwald's step rule. Since ice $I_c$ is kinetically faster to form than ice $I_h$, $I_c$ fragments may grow on the surface of the slowly growing $I_h$ ice clusters. However, ice $I_h$ is thermodynamically more stable, so these $I_c$ fragments can also rapidly melt, leaving the main body of the $I_h$ nuclei to continue to grow. In another study[33], we found that 2-D ice clusters are also subject to a similar process of partial melting before freezing. Further studies are needed to shed more light into this process.

**4 Conclusions**

In summary, we propose here a generalised nucleation theory that describes satisfactorily the kinetics of ice nucleus growth in general conditions. Using all-atom MD simulations, we find that the ice formation from spontaneously grown ice nuclei deviates substantially from the classical nucleation theory, primarily due to defects existing in the ice nuclei. We then proposed to correct the CNT by incorporating the orientation order parameter $q$ as a descriptor of these defects. The kinetics of ice nuclei in various states becomes well described with this simple correction. This result shows that defects within the nucleus are crucial and must be considered in studies of nucleation kinetics. The approach proposed in this letter can also be applied to the kinetics study of other tetra-coordinated crystalline materials. For other types of crystals, alter-native descriptors that characterise internal defects, such as $Q_3$[18,23,34], $Q_4$[19], $Q_6^{18,[19],22,23,[35]}$, can be used. We believe that this generalised nucleation theory has broad theoretical implications for nucleation kinetics and can inspire the study of various biochemistry and material systems[36-38].

**Data availability.** The data supporting the findings of this study are available within the paper, and a detailed description of the calculations is included in the Supplementary Information. All 454 ice nuclei and 180 ideal nuclei generated for the study are available as .xyz files in the Supplementary Information.

**Author Information** The authors declare no competing financial interests. Readers are welcome to comment on the online version of the paper. Correspondence and requests for materials should be addressed to Y.Q.G. (gaoyq@pku.edu.cn) or Y.I.Y. (yangyi@szbl.ac.cn).

**Author Contributions** Conceptualization, Y.Q.G., Y.I.Y. and M.L.; Methodology, Y.I.Y. and M.L.; Software, H.P., Y.X. and C.F.; Investigation, Y.Q.G., Y.I.Y., M.L., H.P., Y.X., D.C., C.F. and L.Y.; Writing – Original Draft Preparation, M.L.; Writing – Review & Editing, Y.Q.G., Y.I.Y.; Funding Acquisition, Y.Q.G., Y.I.Y.

**Notes** The authors declare no competing financial interest.

**Acknowledgements** The authors thank Haiyang Niu, Zhiqiang Ye for useful discussion. Computational resources were supported by the Shenzhen Bay Lab Supercomputing Centre. This research was supported by the National Science and Technology Major Project (No. 2022ZD0115003), the National Natural Science